# Quasi-specular albedo of cold neutrons from powder of nanoparticles

**R. Cubitt[1], E.V. Lychagin[2], A.Yu. Muzychka[2], G.V. Nekhaev[2], V.V. Nesvizhevsky[1*], G. Pignol[3], K.V. Protasov[3], A.V. Strelkov[2]**

[1] *Institut Laue-Langevin, 6 rue Jules Horowitz, Grenoble, France, 38042*
[2] *Joint Institute for Nuclear Research, 6 Joliot-Curie, Dubna, Russia, 141980*
[3] *Laboratoire de Physique Subatomique et de Cosmologie, IN2P3/UJF, 53 rue des Martyrs, Grenoble, France, 38026*

**Abstract**

We predicted and observed for the first time the quasi-specular albedo of cold neutrons at small incidence angles from a powder of nanoparticles. This albedo (reflection) is due to multiple neutron small-angle scattering. The reflection angle as well as the half-width of angular distribution of reflected neutrons is approximately equal to the incidence angle. The measured reflection probability was equal to $\sim 30\%$ within the detector angular size that corresponds to $40-50\%$ total calculated probability of quasi-specular reflection.

---

Coherent scattering of ultracold (UCN), very cold (VCN) and cold (CN) neutrons on nanoparticles could be used (1), (2) in fundamental and applied low-energy neutron physics (3), (4), (5), (6). A theoretical analysis of such scattering could be found, for instance, in (7). In the first Born approximation, the scattering amplitude equals

$$f(\theta)=-\frac{2mU_0}{\hbar^2}r^3\left(\frac{sin(qr)}{(qr)^3}-\frac{cos(qr)}{(qr)^2}\right),\ q=2ksin\left(\frac{\theta}{2}\right) \quad (1)$$

where $\theta$ is the scattering angle, $m$ is the neutron mass, $U_0$ is the real part of the nanoparticle optical potential (8), $\hbar$ is the Planck constant, $r$ is the nanoparticle radius, $k=\frac{2\pi}{\lambda}$ is the neutron wave vector, and $\lambda$ is the neutron wavelength. The scattering cross-section equals

$$\sigma_s=\int|f|^2d\Omega=2\pi\left|\frac{2m}{\hbar^2}U_0\right|r^6\frac{1}{4(kr)^2}\left(1-\frac{1}{(kr)^2}+\frac{sin(2kr)}{(kr)^3}-\frac{sin^2(kr)}{(kr)^4}\right). \quad (2)$$

Consider an idealized case of reflection of a not-decaying neutron from infinitely-thick loss-free powder of nanoparticles occupying half-space. After multiple scattering events the neutron returns to surface. In the case of non-zero imaginary part of the optical potential $U_1$, finite absorption in nanoparticles with the cross-section

$$\sigma_a=\frac{4\pi}{3}\frac{2m}{\hbar^2}U_1r^4\frac{1}{kr} \quad (3)$$

decreases reflectivity. Nevertheless, neutrons with some wave vector are efficiently reflected. Such an albedo of VCN from powder of diamond nanoparticles (9), (10) has been measured (11), (12), providing the best available reflector in a broad energy range. In particular, neutrons with a wavelength $\lambda>2nm$ almost totally reflect from powder of diamond nanoparticles with average radius $\bar{r}\sim 2nm$ at any incident angle. Diamond nanoparticles are chosen for the exceptionally large optical potential of diamond as well as their availability in nearly optimum sizes. The optimum ratio between the neutron wavelength and the nanoparticle size is $\lambda\sim r$ here; then the scattering angle and the cross-section are large (see eqs. 1 and 2).

Now consider faster neutrons so that $\lambda\ll r$ and their finite absorption in nanoparticles. If so, the angle of neutron scattering on each nanoparticle is small (see eq. 1). Therefore neutrons arriving at a large incidence angle penetrate too deep into powder and do not return to the surface until their absorption. Neutrons arriving at a small incidence angle $\alpha$ could return to surface after multiple small-angle scattering. Such neutron albedo is analogous to the process considered in a general form in ref. (13), where an analytic expression describing the angular spectrum of reflected radiation is found for various laws of single scattering of ions, electrons, protons and photons from medium consisting of the scattering centers with sizes significantly larger than the radiation wavelength. As the typical number of scattering events is small, the exit angle $\beta$ is not much higher than $\alpha$, see Fig. 1.

* nesvizhevsky@ill.eu

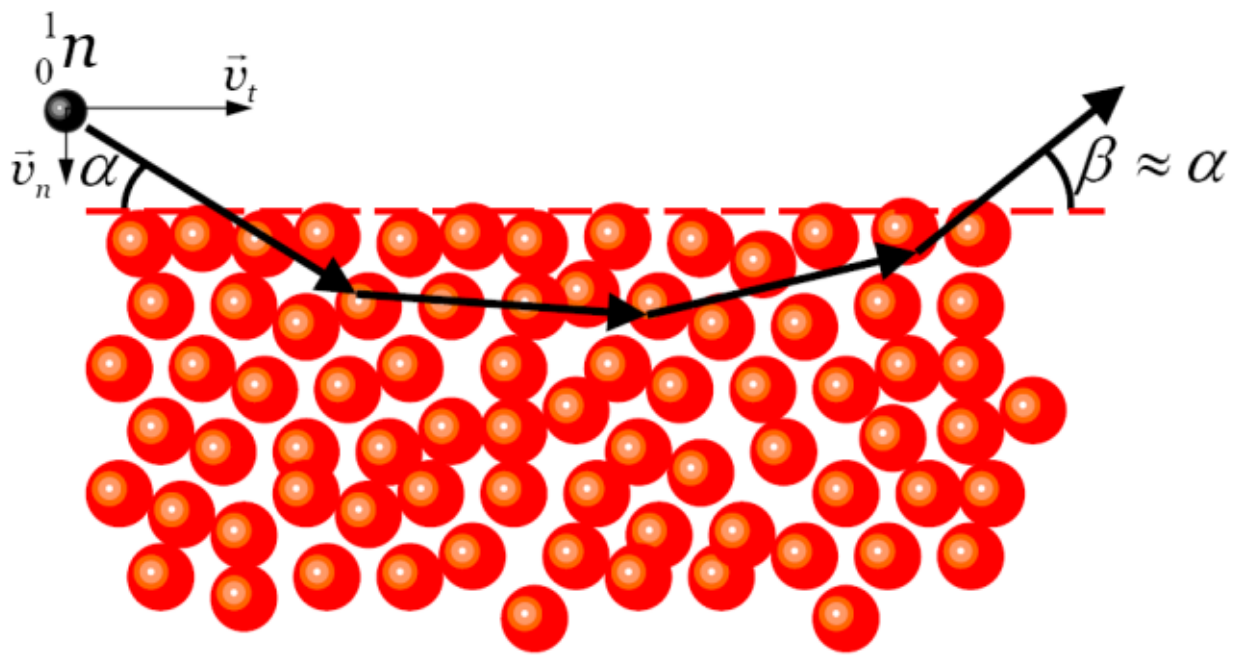

*Fig. 1. Sketch of quasi-specular reflection of a cold neutron from powder of nanoparticles*

In fact, the most probable exit angle $\beta$ is approximately equal to the incidence angle $\alpha$. Such a reflection mechanism is called quasi-specular reflection, or quasi-specular albedo. One should note that the penetration depth and path of neutrons in the powder are relatively small; therefore the absorption affects reflectivity much less than in (11), (12).

The measurements were carried out on the D17 reflectometer (14) at the Institut Laue-Langevin. The measuring scheme (view from above) is shown in Fig. 2. The neutron beam was shaped with two diaphragms. A height and width of a second diaphragm were $15mm$ and $0.3mm$; defining the beam size at the sample. The first diaphragm defined an angular divergence of the beam of $0.0004rad$ in the horizontal plane. The chopper provided time-of-flight neutron spectrum measurements to establish the wavelength. Reflected neutrons were counted in a position-sensitive rectangular neutron detector with a height and width of $50cm$ and $25cm$. Its spatial resolution in height and width was $3mm$ and $2.3mm$. The distance between the sample center and detector was $110cm$. The sample was a powder of diamond nanoparticles with a density of $\sim 0.4g/cm^3$ placed into a prism-shape container with a height of $5cm$, a length of $\sim 15cm$, and a depth of $4cm$. It was inserted into a special cryostat. The surface of the powder on the neutron beam side was covered with $Al$ foil with a thickness of $100\mu m$. Short vertical sides of the prism were covered with $Cd$ plates with a thickness of $0.5mm$. The cryostat allowed for annealing the sample in vacuum at a temperature of $200^o C$, or cooling it down to liquid nitrogen temperature. The sample temperature was measured using a thermo-couple in the middle of the sample. Ballast helium fills in the cryostat and sample for providing thermo-conductivity; thus the sample cooling takes two hours. Neutron scattering from the cryostat window walls was negligible compared to scattering from the sample.

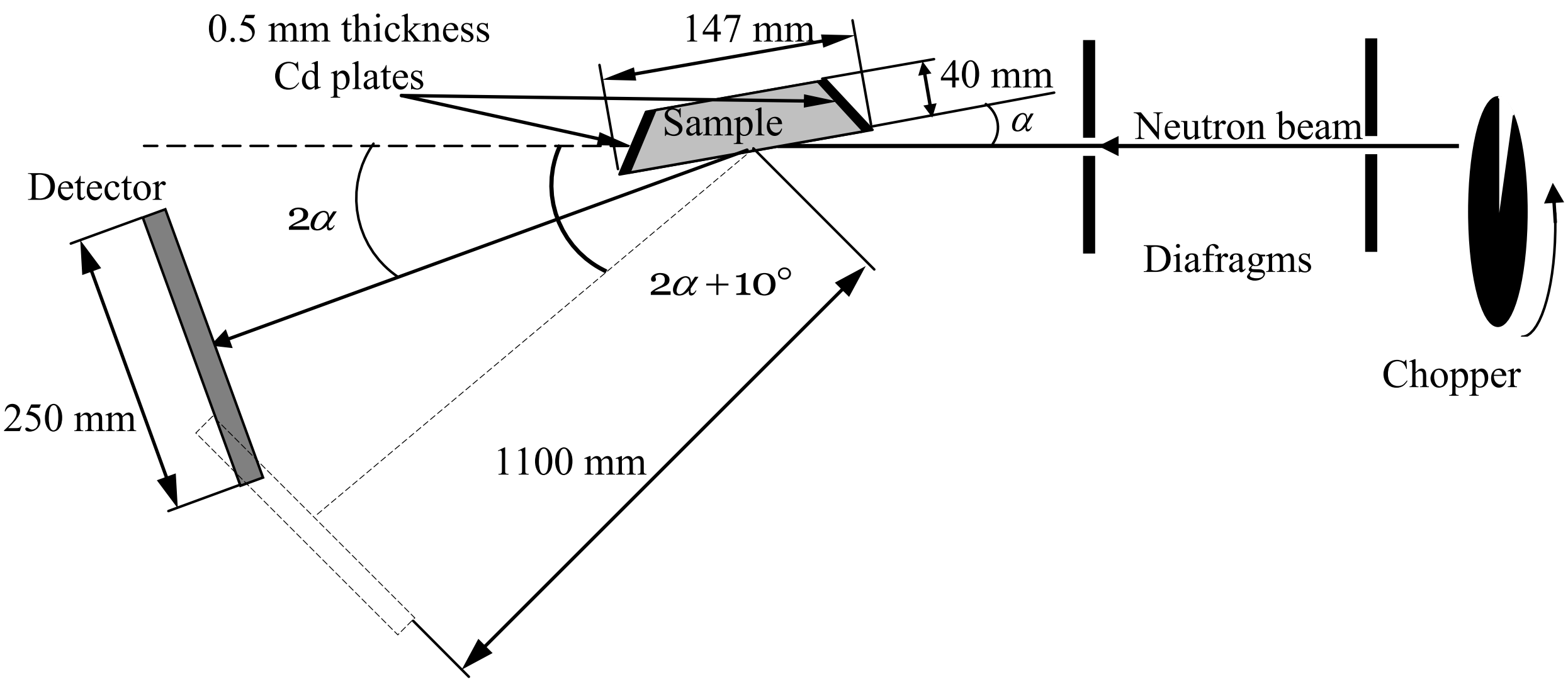

*Fig. 2. Scheme of the experiment (top view)*

We studied the angular dependence of the neutron reflection probability as a function of the neutron wavelength ($2-30$Å) and the incidence angle ($2^o, 3^o$ $and$ $4^o$). The incidence angle $\alpha'$ equals zero when neutron beam was parallel to the sample surface; the vertical rotation axis of the sample table crosses the sample center. The spectrum and flux of incident neutrons was measured with the sample shifted horizontally by $\sim 1cm$ out of the beam, the detector was rotated to an angle $2\alpha = 0^o$ and shifted horizontally by $\sim 1.5cm$. In the scattering measurements, the sample was rotated by an angle $\alpha$ and the detector was rotated by an angle $2\alpha$. For each incidence angle $\alpha$, the sample was shifted horizontally perpendicular to the beam axis in order to maximize the flux of neutrons scattered to the detector (this translation was equal to $0.0; 0.5; 1.0mm$ respectively). For each grazing angle $\alpha$, the detector was rotated also to an angle $2\alpha + 10^o$. Thus, we measured two-dimensional distributions of scattered neutrons within the azimuth angle of $24^o$ and the polar angle range between $0^o$ and $\sim \alpha + 15^o$.

All measurements were carried out twice: 1) at ambient temperature, after preceding sample treatment (long annealing and heating at $150^o C$); 2) at liquid nitrogen temperature. The total measuring time was 14 hours.

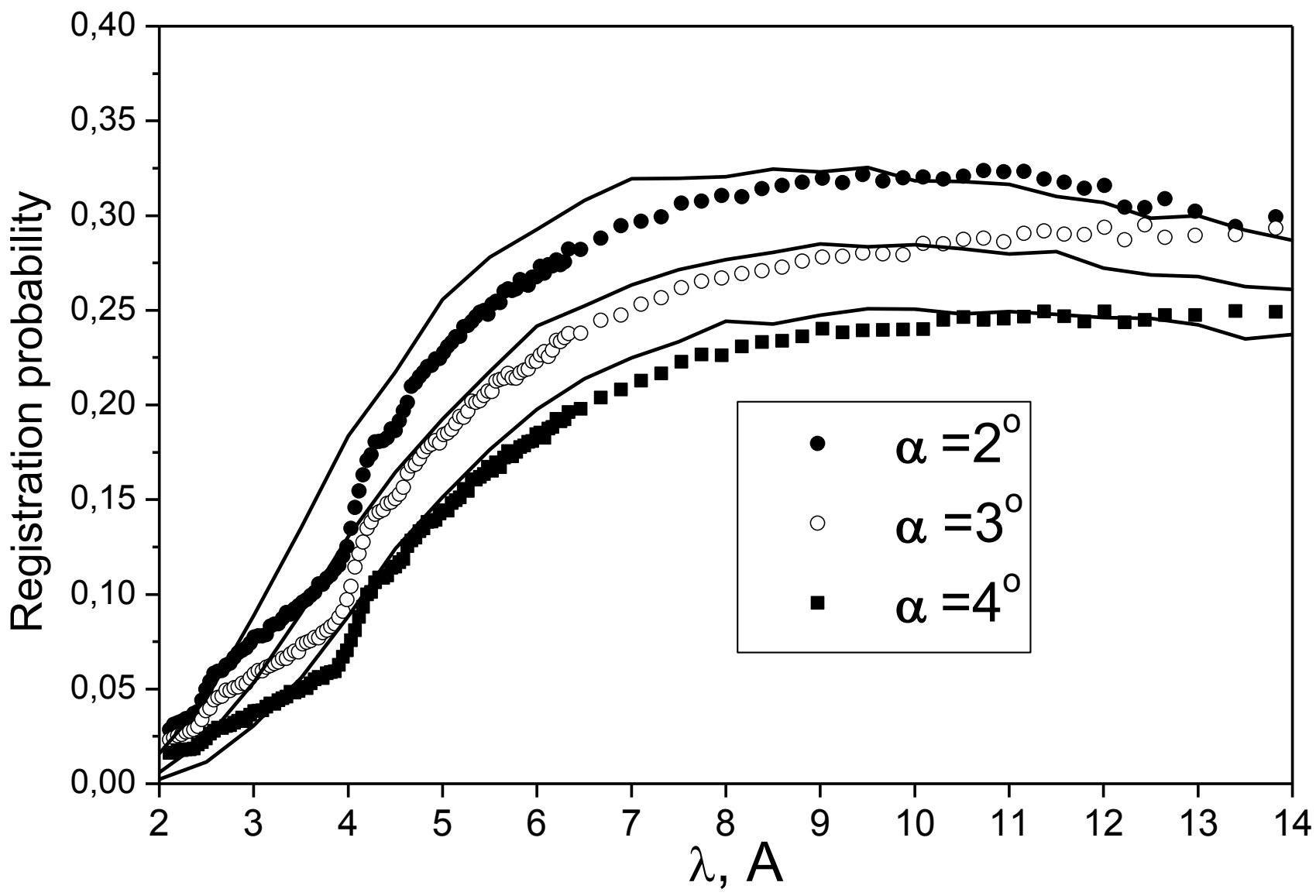

*Fig. 3. The probability of neutron reflection within the detector solid angle is shown as a function of the neutron wavelength. The incidence angle α is equal to* $2^o$, $3^o$ and $4^o$. *Dark and empty circles as well as squares correspond to measured data; solid lines illustrate calculations.*

Fig. 3 shows the probability of neutron reflection within the detector solid angle as a function of the neutron wavelength and the incidence angle (as defined in Fig. 2). The reflectivity values in Fig. 3 are smaller than actual ones by a fraction of neutrons scattered to angles larger than the detector solid angle. Results of measurements at ambient and nitrogen temperature do not differ significantly. In particular, this is due to the small number of scattering events involved into quasi-specular reflection. Besides, the temperature-dependent inelastic neutron scattering is small ($\sigma_{in}(2200m/s) = 1b$) compared to temperature-dependent elastic cross-section ($\sigma_{el} = 120b$), and a fraction of hydrogen atoms in annealed nanoparticles is low: $1/15$. Moreover, hydrogen is strongly bound to carbon; the phonon excitation spectrum is close to that for diamond. Any neutron scattering at hydrogen (both elastic and inelastic ones) is isotropic, therefore such a scattered neutron is almost totally lost. We estimate neutron losses at hydrogen to be equal to $20-40\%$ for various incident angles $\alpha$ actually used. The wavelength range of effective quasi-specular reflection is limited to below $\sim 4$Å by Bragg scattering of neutrons in the bulk of a diamond nanoparticle.

Computer simulation of quasi-specular reflection is straightforward; it is based on formulas (1-3). The simulation method has been verified in our preceding works (11), (12); it is useful for understanding general features of the phenomenon. The measurement geometry used here is shown in Fig. 2. The powder density, a fraction of hydrogen and the scattering cross-sections are given above. The measured data are compared to the model for the neutron incidence angle $2^o$ and wavelength $\lambda = 10Å$. Within a simplified hypothesis of equal size distribution of nanoparticles, such an average diameter appeared to be equal to $2nm$. This model was expanded to other values of the neutron incidence angles and wavelengths. The calculated reflection probabilities defined as described above are shown in Fig. 3 with solid lines.

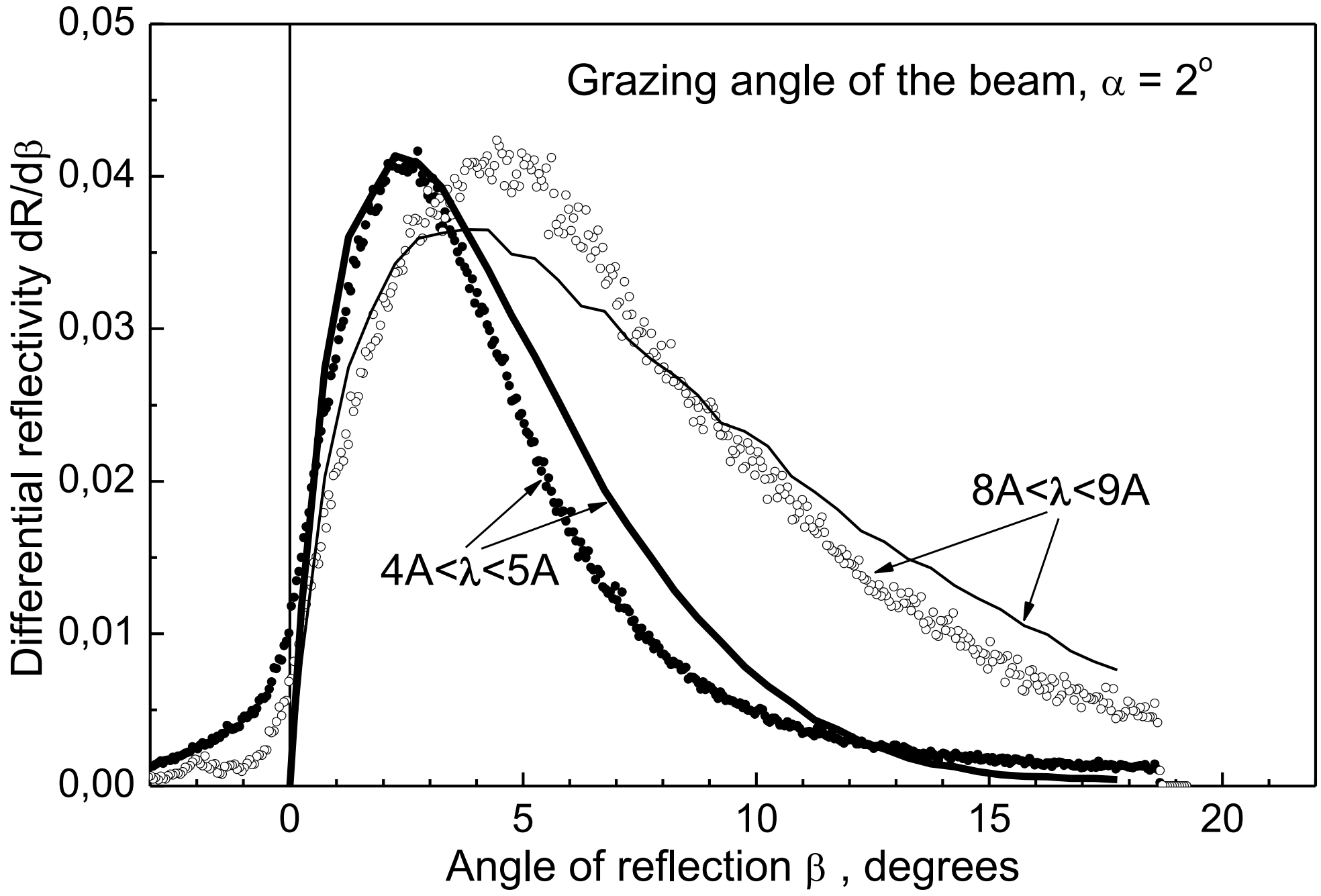

*Fig. 4. Angular distributions of reflected neutrons. The incidence angle is equal* $2^o$*.*
*Dark and empty circles correspond to measured data; solid lines illustrate calculations.*

Fig. 4 shows measured (dark and empty circles) and calculated (solid lines) angular distributions of reflected neutrons; the neutron incidence angle is $2^o$. The data are averaged over two ranges of wavelengths of incident neutrons: $4-5Å, and\ 8-9Å$. The neutron count rates in the position-sensitive detector are integrated over vertical stripes and divided by the incident neutron flux at the same wavelengths. Analogous wavelength and angular distributions were measured for incidence angles $3^o$ and $4^o$. Fig. 4 indicates that some neutrons reach the detector at scattering angles smaller than $0^o$ that contradicts the geometrical constraint shown in Fig. 2. This is probably due to some bloating of the sample during its annealing; the surface deflected by $\sim 2mm$. However, we will ignore here this correction as we are interested in the observation of a new physical phenomenon and in the investigations of its general features rather than in its precision analysis at this stage of our research. The angular distributions of reflected neutrons are calculated within the model described above. Some broadening of the calculated angular distributions compared to the measured data is explained by the simplification of the model (equal sizes of nanoparticles). Nevertheless, the general agreement of the data and such a simple model are quite good.

To conclude, we predicted and observed for the first time the phenomenon of quasi-specular albedo, at small incidence angles, of cold neutrons from powder of nanoparticles. For the diamond nanoparticles used, the wavelength range of effective quasi-specular reflection is limited from below at $\sim 4Å$ by Bragg scattering in diamond. Quasi-specular reflection could find numerous applications, in particular, as neutron reflectors in zones close to the reactor core, where other types of reflectors would not survive radiation damage. Such reflectors would increase significantly the flux of cold neutrons available for experiments.

We are grateful to Federal Agency of Education (Rosobrazovanie), Russia, grant НК-20П(3), for financial support.

## References

1. **Nesvizhevsky, V.V.** Interaction of neutrons with nanoparticles. *Physics of Atomic Nuclei.* 2002, Vol. 65(3), 400.
2. **Artemiev, V.A.** Estimation of neutron reflection from nanodispersed materials. *Atomic Energy.* 2006, Vol. 101, 901.
3. **Ignatovich, V.K.** *The Physics of Ultracold Neutrons.* Oxford : Clarendon Press, 1990.
4. **Golub, R., Richardson, D.J., and Lamoreux, S.K.** *Ultracold Neutrons.* Bristol : Higler, 1991.
5. **Pendlebury, J.M.** Fundamental physics with ultracold neutrons. *Annual Review of Nuclear and Particle Science.* 1993, Vol. 43, 687.
6. **Abele, H.** The neutron. Its properties and basic interactions. *Progress in Particle and Nuclear Physics.* 2008, Vol. 60, 1.
7. **Nesvizhevsky, V.V., Pignol, G., and Protasov, K.V.** Nanoparticles as a possible moderator for an ultracold neutron source. *International Journal of Nanoscience.* 2007, Vol. 6(6), 485.
8. **Fermi, E., and Marshall, L.** Interference phenomena of slow neutrons. *Physical Review.* 1947, Vol. 71, 666.
9. **de Carli, P.J., and Jameieson, J.C.** Formation of diamond by explosive shock. *Science.* 1961, Vol. 133, 1821.
10. **Aleksenskii, A.E., Baidakova, M.V., Vul', A.Y., and Siklitskii, V.I.** The structure of diamond nanostructures. *Physics of Solid State.* 1999, Vol. 41, 668.
11. **Nesvizhevsky, V.V., Lychagin, E.V., Muzychka, A.Yu., Strelkov, A.V., Pignol, G., and Protasov, K.V.** The reflection of very cold neutrons from diamond powder nanoparticles. *Nuclear Instruments and Methods A.* 2008, Vol. 595(3), 631.
12. **Lychagin, E.V., Muzychka, A.Yu., Nesvizhevsky, V.V., Pignol, G., Protasov, K.V., and Strelkov, A.V.** Storage of very cold neutrons in a trap with nano-structured walls. *Physics Letters B.* 2009, Vol. 679, 186.
13. **Remizovich, V.S.** Theoretical description of elastic reflection of particles (photons) incident at grazing angles without the use of the diffusion approximation. *JETP.* 1984, Vol. 60(2), 290.
14. **Cubitt, R., and Fragneto, G.** D17: the new reflectometer at the ILL. *Applied Physics A.* 2002, Vol. 74, S329.